\documentclass{article}
\usepackage{amsmath,amssymb}
\begin{document}

\title{~\\~\\New integrable systems\\and\\ a curious realisation of $SO(N)$}

\author{{\bf Jens Hoppe}\footnote{Email address: hoppe@math.kth.se}\\~\\}

\maketitle

\begin{center}
\vskip 8mm {\small\it Department of Mathematics \\ Royal Institute
of Technology, Lindstedtsv\"agen 25 S-10044 Stockholm, Sweden \\
and \\ Center for Theoretical Physics
\\Seoul National University, 151-742 Seoul, Korea}
 \vskip 10mm
\par

\begin{abstract}
A multiparameter class of integrable systems is introduced.
\end{abstract}

\end{center}

\vfill


\newpage





\newcommand{\be}{\begin{equation}}
\newcommand{\ee}{\end{equation}}
\newcommand{\ba}{\begin{eqnarray}}
\newcommand{\ea}{\end{eqnarray}}
\def\d{\delta}
\def\o{\omega}
\def\s{\sigma}
\def\p{\partial}
\def\a{{\alpha}}
\def\ra{\rangle}
\def\la{\langle}
\def\rg{\ra}
\def\lg{\la}

%

\setcounter{equation}{0}

As is well known, geodesic motions on a manifold are integrable
only in very special cases. Jacobi's ingenious solution for
ellipsoids\footnote{announced 168 years ago in a letter to the
French Academy \cite{Jacobi1}.} and the impact of his invention of
elliptical coordinates, explained in detail in his K\"onigsberg
lectures \cite{Jacobi2}, can hardly be underestimated (reading
through his five page note to the Prussian Academy \cite{Jacobi3},
one feels that, despite so much progress during the century/ies
that followed, (t)his kind of art is gone forever).

Comparatively recently (\cite{4}, see also \cite{5} - \cite{7})
the commutativity of the $N$ quantities
\be\label{F} F_i \equiv p_i^2 + {\sum^N_{j=1}}  {}^{'}\frac{(x_i
p_j - x_j
 p_i)^2}{\alpha_i - \alpha_j} \ee
for the ellipsoid ($\sum^N_{i=1} \frac{x_i^2}{\a_i}=1$), resp.
\be\label{G} G_i \equiv x_i^2 + {\sum^N_{j=1}}  {}^{'}\frac{(x_i
p_j - x_j
 p_i)^2}{\alpha_i - \alpha_j} \ee
for the related Neumann-problem \cite{8} of $N$ particles moving
on the sphere $(\sum^N_{i=1} x_i^2 =1$) subject to the external
potential $V(\vec x) = \frac12 \sum \a_i x_i^2$, was noticed.

The commutativity with respect to
\be\label{Poisson} \{f,g\}(\vec x, \vec p) \equiv \sum_{i=1}^N
\left(\frac{\p f}{\p x_i}\frac{\p g}{\p p_i} - \frac{\p f}{\p
p_i}\frac{\p g}{\p x_i}\right) \ee
follows straightforwardly from
\be\label{cyclic} \frac1{\a_{ik}\a_{kl}} + \frac1{\a_{kl}\a_{li}}
+ \frac1{\a_{li}\a_{ik}} = 0 \ee
($\a_{ij}\equiv \a_i -\a_j$), and the standard angular momentum
algebra relations,
\be\label{JJ} [J_{ij}, J_{kl}]= -\d_{jk} J_{il} +\d_{ik} J_{jl}
+\d_{jl} J_{ik} -\d_{il} J_{jk} \ee
\be\label{JxJp} \begin{matrix} [J_{ij}, x_k] = \d_{ik} x_j - \d_{jk} x_i \\
[J_{ij}, p_k] = \d_{ik} p_j - \d_{jk} p_i \end{matrix} \ee
satisfied by $J_{ij} \equiv x_i p_j - x_j p_i$ (and
$[~,~]\equiv\{~,~\}$).

In particular,
\begin{eqnarray}  &~& \frac14\left[\sum_j {}^{'}
\frac{L_{ij}^2}{\a_{ij}},\sum_l {}^{'}
\frac{L_{kl}^2}{\a_{kl}}\right] \nonumber  \\\label{bracket1} &~&
~~~~ = \sum_{j,l} {}^{''}
\frac{L_{ij}L_{kl}}{\a_{ij}\a_{kl}}(-\d_{jk}L_{il}+\d_{ik}L_{jl}+\d_{jl}L_{ik}
- \d_{il}L_{jk}) \\  &~& ~~~~ =  \d_{ik} \left(\sum_{j,l} {}^{''}
\frac{L_{ij}L_{il}L_{jl}}{\a_{ij}\a_{il}}\right) - L_{ik}\sum_l
{}^{''} L_{il}
L_{kl}\left(\frac1{\a_{il}\a_{lk}}+\frac1{\a_{lk}\a_{ki}}+\frac1{\a_{ki}\a_{il}}\right)
\nonumber
\end{eqnarray}
is equal to zero as long as the $L_{ij}$ satisfy (\ref{JJ}).

The commutativity with respect to the constrained (Dirac) bracket,
\be\label{dirac} \{f,g\}_{\rm D} \equiv \{f,g\} +
\{f,\varphi\}\frac1{J}\{\Pi, g\} - \{f,\Pi\}\frac1{J}\{\varphi,
g\}~,\ee
with $\varphi = \frac12(\vec x^2 - 1)$, $\Pi = \vec x\cdot \vec
p$, ($J\equiv \{\varphi, \Pi\} = 1$) for the Neumann-problem, and
$\varphi = \frac12(\sum_{i=1}^N \frac{x_i^2}{\a_i} - 1)$, $\Pi =
\sum_{i=1}^N \frac{x_i p_i }{\a_i}$, ($J\equiv \{\varphi, \Pi\}$)
for the $N$-dimensional ellipsoid, then follows by noting the
respective commutativity of $\varphi$ with the $F_i$, resp. $G_i$.

Let me point out that (\ref{JJ}) and (\ref{JxJp}) also imply the
commutativity of the quantities
\be\label{5plus6} J_i = \a x_i p_i + \sum_j {}^{'} \frac{(x_i p_j
- x_j
 p_i)^2}{\alpha_i - \alpha_j}~,\ee
as for $i\neq k$ (and anticipating the quantum-commutativity by
keeping track of the order)
\be \label{xpJ} \left[x_k p_k, \sum_j
{}^{'}\frac{J_{ij}^2}{\a_{ij}}\right] = x_k \frac{J_{ik}}{\a_{ik}}
p_i + x_i \frac{J_{ik}}{\a_{ik}} p_k + x_k p_i
\frac{J_{ik}}{\a_{ik}} +  \frac{J_{ik}}{\a_{ik}}x_i p_k \ee
is symmetric under the interchange of $i$ and $k$, as well as note
the formal commutativity of the differential operators
($k=1,\cdots,N$)
\be\label{Hk} \hat H_k = -\sum_l {}^{'} (x_k x_l)^{1/4}(\p_k
-\p_l)\frac{\sqrt{x_k x_l}}{\a_{kl}}(\p_k -\p_l)(x_k x_l)^{1/4} -
\frac{i\a}2 (x_k \p_k +\p_k x_k)\ee
(acting on functions of $\vec x \in {\bf R}_+^N$, $\a \in {\bf R}$
and arbitrary $\a_{kl}=-\a_{lk}\neq 0$, satisfying
(\ref{cyclic})). The operators (with their classical counterparts
$L_{ij}\equiv -2\sqrt{x_i x_j} (p_i-p_j)$)
\be \label{L} \hat L_{ij} \equiv 2(x_i x_j)^{1/4}
\left(\frac{\p}{\p x_i} - \frac{\p}{\p x_j}\right) (x_i x_j)^{1/4}
\equiv 2 \rho_{ij} \p_{ij} \rho_{ij} \ee
satisfy the $so(N)$ Lie-algebra relations (\ref{JJ}) (from now on,
$[~,~]$ denoting the ordinary commutator), as well as
\be \label{xpL} [x_k \p_k, \hat L_{ij}] = (\d_{kj}-\d_{ki})
\left((\p_i+\p_j) \rho_{ij}^2 + \rho_{ij}^2
(\p_i+\p_j)\right)~,\ee
\be \begin{matrix} [\hat L_{ij}, x_k] = 2\rho_{ij}^2 (\d_{ik} -
\d_{jk}) \\ [\p_k, \hat L_{ij}] = \frac14
(\d_{ki}+\d_{kj})\left(\frac1{x_k}\hat L_{ij} + \hat L_{ij}
\frac1{x_k}\right) \end{matrix} \ee
\be \label{Lrho} [\hat L_{ij}, \rho_{kl}^2] = \d_{ik} \rho_{jl}^2
\pm 3 ~\text{more}~.\ee
The commuting classical quantities corresponding to (\ref{Hk}) are
($k=1,\cdots, N$)
\be \label{cHk1} \tilde H_k  = \sum_{l\neq k} \frac{x_k
x_l}{\a_{kl}}(p_k -p_l)^2 + \a x_k p_k ~,\ee
resp. (interchanging $\vec x$ and $\vec p$).
\be \label{cHk2} H_k = \sum_{l\neq k} p_k \frac{(x_k -
x_l)^2}{\a_{kl}} p_l  - \a x_k p_k ~.\ee

Arbitrary functions of the commuting quantities (\ref{cHk2})
(resp. (\ref{cHk1})/(\ref{Hk}) or (\ref{5plus6})) can be taken as
Hamiltonians.

Let me finish with some remarks:

\begin{itemize}

\item[-] While ($x_{ij} \equiv x_i -x_j$)
\be \label{xxx} \left[x_k \p_k +\p_k x_k, \sum_j {}^{'} x_{ij}
\frac{\p_i\p_j}{\a_{ij}} x_{ij}\right] - (i\leftrightarrow k) = 0
~,\ee
the naive quantisations of (\ref{cHk2})$_{\a=0}$ do {\it not}
commute:
\ba \label{xxxx}&~& \left[\sum_j{}^{'}
x_{ij}\frac{\p_{ij}^2}{\a_{ij}}x_{ij}, \sum_l {}^{'} x_{kl}
\frac{\p_{kl}^2}{\a_{kl}}x_{kl}\right] \\ &~&~~ =
\frac{x_{ik}}{\a_{il}\a_{kl}} (\p_i+\p_k -\p_l) +
\frac{x_{kl}}{\a_{ki}\a_{li}} (\p_k+\p_l -\p_i) +
\frac{x_{li}}{\a_{lk}\a_{ik}} (\p_l+\p_i -\p_k)~.\nonumber\ea

Any other (hermitian) ordering of the 4 quantities $x_{ij}$,
$\p_i$, $\p_j$, $\p_{ij}$ gives the same result (This discrepancy
compared to the commutativity of (\ref{Hk}), is due to having
singled out the coordinate-representation, resp. avoiding
$\sqrt{\p_i\p_j}$.).

\item[-] For $N=2$, ($\a=0$) and the choice $H=\sum_k
\frac{H_k}{\a_k} = \frac12 \sum_{k,l} p_k
\frac{x_{kl}^2}{\a_k\a_l} p_l$, e.g., one would get (with $q\equiv
x_2 - x_1$, $p\equiv \frac{p_2-p_1}2$, $P\equiv p_1+p_2 =
\rm{const}$., $\mu\equiv \a_1\a_2$),
\be \label{quart} H = \frac1{2\mu} \left(\frac{P^2}4 q^2 - p^2
q^2\right)~,\ee
\be\label{eom} \ddot q = \frac{\dot q^2}{q} +
\frac{P^2}{4\mu^2}q^3~,\ee
resp.
\be\label{E} \dot q^2 = \frac{P^2}{4\mu^2} q^4 -\frac{2E}{\mu}
q^2~.\ee
The integration is elementary.

\item[-] Apart from questions of domains, (\ref{Hk}) is equivalent
to the quantisation of (\ref{5plus6}).

\end{itemize}

\vspace{1cm}

\begin{center}
{\bf Acknowledgement}
\end{center}

\vspace{0.5cm}

\noindent I would like to thank Joakim Arnlind, Martin Bordemann,
Min-Young Choi, Choonkyu Lee, and Kimyeong Lee for discussions, as
well as the Swedish Science Foundation, the Brainpool program of
the Korea Research Foundation and the Korean Federation of Science
and Technology Societies, R14-2003-012-01002-0, and the Marie
Curie Training Network ENIGMA, for support.


\begin{thebibliography}{99}
\bibitem{Jacobi1} C.G.J.~Jacobi, {\em Lettre de M. Jacobi \`a M. Arago
concernant les lignes g\'eod\'esiques trac\'ees sur un ellipsoide
\`a trois axes}, Compt. Rend. {\bf 8}, 284 (1839).

\bibitem{Jacobi2} C.G.J.~Jacobi, {\em Vorlesungen \"uber Dynamik} (which
Jacobi gave in 1842/43 at K\"onigsberg University) Reimer 1866.

\bibitem{Jacobi3} C.G.J.~Jacobi, {\em Note von der geod\"atischen Linie
auf einem Ellipsoid und den verschiedenen Anwendungen einer
merkw\"urdigen analytischen Substitution}, JRAM {\bf 19}, 309--313
(1839).

\bibitem{4} K.~Uhlenbeck, {\em Equivariant harmonic maps into spheres}.
\newblock Springer Lecture Notes in Mathematics {\bf 949} (1982), p. 146.

\bibitem{5} J.~Moser.
\newblock Various aspects of hamiltonian systems.
\newblock In {\em Proceedings of the C.I.M.E. Bressanone. Progress in Math.}, volume~8. Birkh\"auser,  (1978).

\bibitem{6} O.~Babelon and M.~Talon,
Nucl.\ Phys.\ B {\bf 379}, 321 (1992) [arXiv:hep-th/9201035].

\bibitem{7} A.~Perelomov,
Regular and Chaotic Dynamics 5, No. 1, 89-91 (2000)
[arXiv:math-ph/0203032].

\bibitem{8} C.~Neumann, {\em De problemate quodam mechanico, quod ad primam integralium
ultraellipticorum classem revocatur}.
\newblock JRAM {\bf 56} (1859), p.~46.

\end{thebibliography}
\end{document}